\documentclass[aps,prd,floatfix,showpacs]{revtex4}
\usepackage[dvips]{graphicx}
\newcommand{\be}{\begin{equation}}
\newcommand{\ee}{\end{equation}}
\newcommand{\bea}{\begin{eqnarray}}
\newcommand{\eea}{\end{eqnarray}}
\newcommand{\bitem}{\begin{itemize}}
\newcommand{\eitem}{\end{itemize}}

\newcommand{\benum}{\begin{enumerate}}
\newcommand{\eenum}{\end{enumerate}}
\newcommand{\bc}{\begin{center}}
\newcommand{\ec}{\end{center}}
\begin{document}
\title{SCALE INVARIANT STELLAR STRUCTURE }
\author{Sidney Bludman}
\email{sbludman@das.uchile.cl}
\homepage{http://www.das.uchile.cl/~sbludman}
\affiliation{Departamento de Astronom\'ia, Universidad de Chile, Santiago, Chile}
\author{Dallas C. Kennedy}
\email{dkennedy@mathworks.com}
\homepage{http://home.earthlink.net/~dckennedy}
\affiliation{The MathWorks, Inc., 3 Apple Hill Drive, Natick, Massachusetts, USA}
\date{\today}

\begin{abstract}
In scale invariant hydrostatic barotropes, the radial evolutionary equation linearly relates the local
gravitational and internal energies.  From this first-order equation, directly follow all the properties
of polytropes and the important mass-radius relation. Quadrature then leads to the regular Lane-Emden
functions and their Picard and Pade approximations, which are useful wherever stars are approximately or exactly polytropic. We illustrate this particularly for the n=3 regular polytrope and obtain analytic approximations
to the solution of the Lane-Emden equation, valid over the bulk of relativistic degenerate stars
(massive white dwarfs) and chemically homogeneous stars in radiative equilibrium (ZAMS stars).
\end{abstract}

\maketitle
\tableofcontents

\section{DIFFERENT INDEX POLYTROPES SHOW DIFFERENT CORE CONCENTRATIONS \& MASS-RADIUS RELATIONS} 

In the preceding paper~\cite{BludKenI}, we derived the characteristic differential equations
\be \frac{d\log{v_n}}{u-1+v_n}=\frac{d\log{u}}{3-u -n v_n}=d\log{r} \label{eq:chareqns} \ee
for the homology invariants $u:=d\log{m}/d\log{r},~v_n:=-d\log{P/\rho}/d\log{r}$ of n-th order polytropes. Each
{\em regular polytrope} is
a solution $v_n(u)$ of the first-order Abel equation, subject to the regularity condition $u(0)=3, v_n(0)=0$
 \be \frac{dv_n}{du}=\frac{v_n(u-1+v_n)}{u(3-u-n v_n)}\quad ,\ee
characterized by the boundary value $_0\omega_n:=\omega_n (\xi_{1n})$ of its homology invariant
\be \omega_n  (\xi):=(uv_n^n)^{\tilde{\omega_n}/2}\quad .\ee
As tabulated in the third, fourth and last columns of Table~I~\cite{Chandra,Chui,Schwarzschild,Hansen,Kippen},
$_0\omega_n$ determines the boundary $\xi_{1n}$ and the
density ratio $\rho_{cn}/\bar\rho_n$, where $\bar\rho_n=3M/4\pi R^3$ is the mean density, the $M$-$R$ relation $M\sim~_0\omega_n R^{(n-3)/(n-1)}$. These properties of polytropes result \emph{directly} from the first-order equation (2), which encapsulates the effect of scale invariance, and whose solution for each polytropic index $n$ is plotted in Figures~1 and~2.

\subsection{Regular Polytropes of Different Index Have the Same Inner Core But Different Envelopes} 

After solving the second-order Lane-Emden equation, each Lane-Emden function of index $n$
is also characterized by the first zero of $\theta_n (\xi)$, the dimensionless radius $\xi_{1n}$.
The sixth and
seventh columns in Table~I list dimensionless values for the {\em inner core} radius $r_{{\rm ic}n}/R=\xi_{{\rm ic}n}$ and included mass $m_{{\rm ic}n}/M$, shown by red dots in Figures~3,~4.
The {\em inner core radius} $\xi_{{\rm ic}n}$, defined implicitly by $u(\xi_{{\rm ic}n}):=2$, is where
the acceleration $Gm/r^2$ reaches a maximum and the gravitational energy density overtakes the internal energy density. In the inner core, the internal energy dominates; in the envelope outside,
the gravitational energy dominates.

\begin{table*}[b] 
\caption{Scaling Exponents, Core Parameters, Surface Parameters, and Mass-Radius Relations for Polytropic Gas Spheres
of Increasing Core Concentration}
\begin{ruledtabular}
\begin{tabular}{|l|l||l|l|l||l|r||r|}
$n$ &$\tilde{\omega}_n$ &$\xi_{1n}$ &$\rho_{cn}/\bar\rho_n$&$_0\omega_n$ &$r_{{\rm ic}n}/R$&$m_{{\rm ic}n}/M$
&$M\sim R^{1-\tilde{\omega}_n}$ Properties  \\
\hline 
0   &-2         &2.449              &1                  &0.333          &1         &1            &$R\sim M^{1/3}$; incompressible matter, all core\\
1   &$\pm\infty$&3.142              &3.290              &...            &0.66      &0.60         &$R$ independent of $M$ \\
1.5 &4          &3.654              &5.991              &132.4          &0.55      &0.51         &$R\sim m^{-1/3}$; nonrelativistic degenerate\\
2   &2          &4.353              &11.403             &10.50          &0.41      &0.41         &                     \\
3   &1          &6.897              &54.183             &2.018          &0.24      &0.31         &$R$ independent of $M$; Eddington standard model\\
4   &2/3        &14.972             &622.408            &0.729          &0.13      &0.24         &                      \\
4.5 &4/7        &31.836             &6189.47            &0.394          &0.08      &0.22         &                 \\
5   &1/2        &$\infty$           &$\infty$           &0              &0         &0.19          &maximally compressible, all envelope; $R=\infty$ for any $M$ \\
\end{tabular}
\end{ruledtabular}
\end{table*}

For homology variables, we prefer a new independent variable $z:=3-u=-d\log{\bar{\rho}_n}/d\log{r}$ and a new dependent
variable
$w_n (z):= nv_n:=-d\log{\rho}/d\log{r}$. In term of these invariants, the
characteristic differential equations (1) are
\be \frac{d z}{(3-z)(w_n-z)}=\frac{d\log{w_n}}{2-z+\frac{w_n}{n}}=d \log{r}\quad .\ee
The first equality is the first-order equation (2)  for the invariant $w_n(z)$, which we solve for the central boundary
condition $w_n \rightarrow 5 z/3$ when $z\rightarrow 0$. $w_n(z)$ and the differences $(3/5)[w_n(z)-w_5(z)]=
(3/5)w_n (z)-z=d \log{ \bar{\rho}/\rho^{\frac{3}{5} }} \sim G_n$ are plotted in Figures~2 and~1, respectively,
for polytropic indices $n=1,2,3,4,5$.

\begin{figure}[t] 
\includegraphics[scale=0.70]{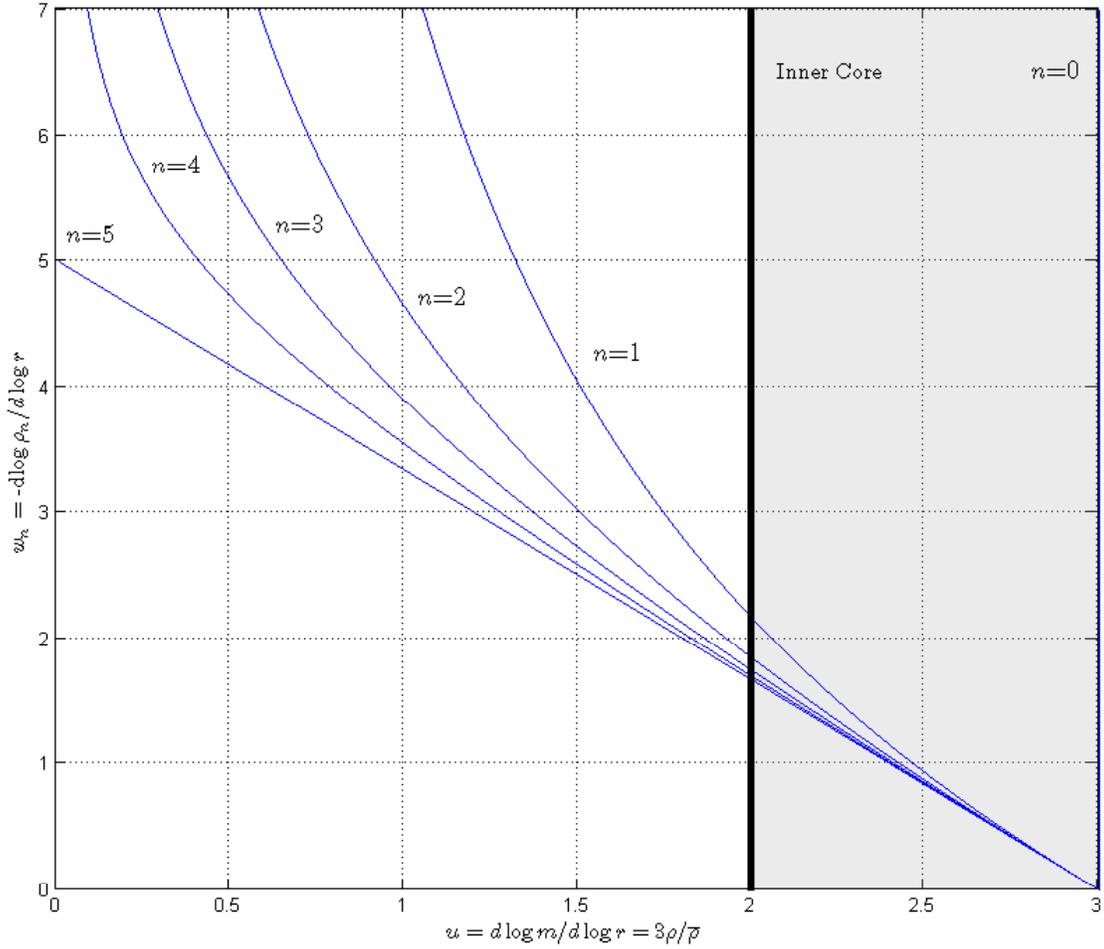}
\caption{Dilution of polytrope density profiles in the envelope as the boundary is approached ($u\rightarrow 0$). All solutions approach the same density structure $w_n(z)\rightarrow w_5=(5/3)(3-u)$ at the center ($u\rightarrow 3$), but differ outside the core. As the boundary is approached, the mass $m\rightarrow M$, and $w_n \rightarrow n[_0\omega_n ^{n-1}/u]^{1/n}$,
$r/w_n\rightarrow 0$.}
\end{figure}

 \begin{figure}[t] 
\includegraphics[scale=0.70]{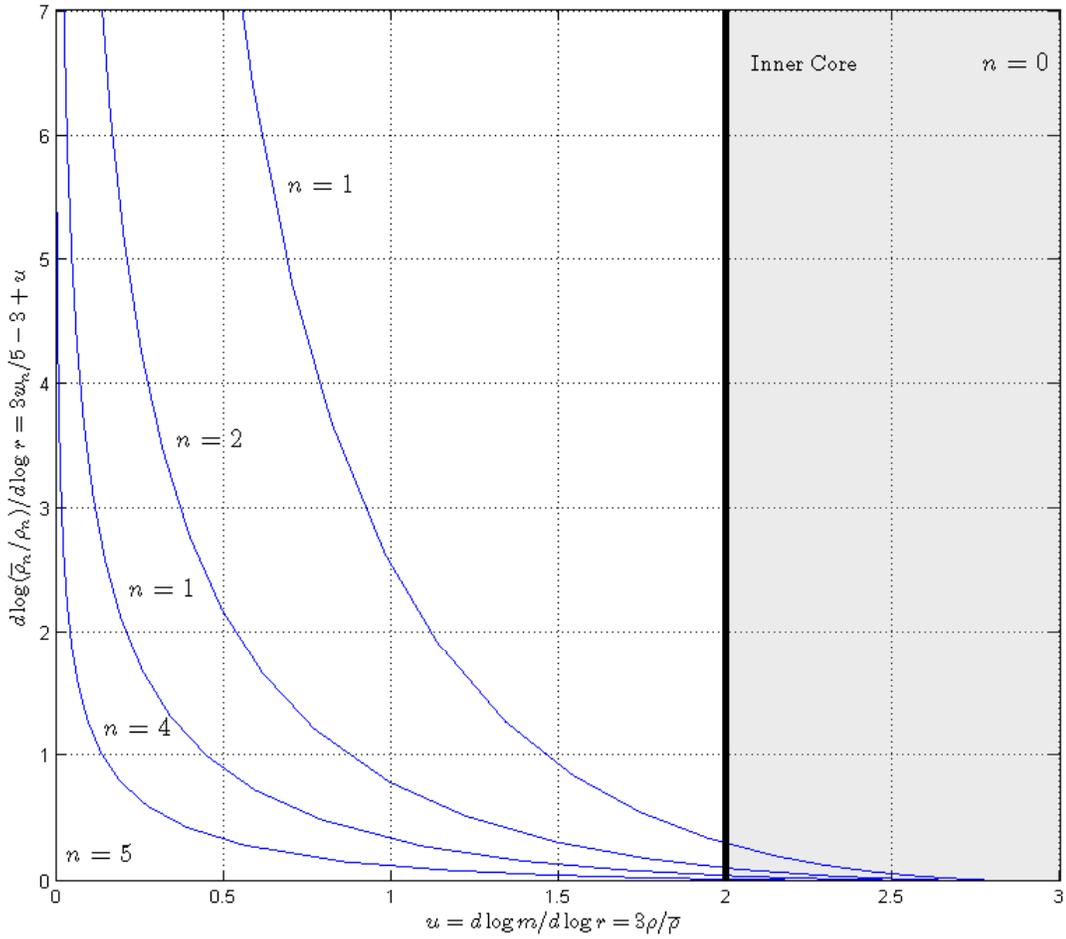}
\caption
{Different Lane-Emden density gradients $w_n (u):=-d\log{\rho}/d\log{r}$ as the outer boundary $u=0$ is approached. Inside the core $u<2$, all $w_n\approx w_5=(5/3) (3-u)$.
For the softest equation of state $n=5$,
the stellar radius is infinite, the density gradient $w_5(u)=(5/3)(3-u)$ everywhere, and $\bar{\rho}=\rho_c ^{2/5} \rho ^{3/5}$. For stiffer equations of state $n<5$, the density gradients $w_n (u)$ steepen as the finite radius is approached $u \rightarrow 0$.}
\end{figure}

\begin{figure}[t] 
\includegraphics[scale=0.65]{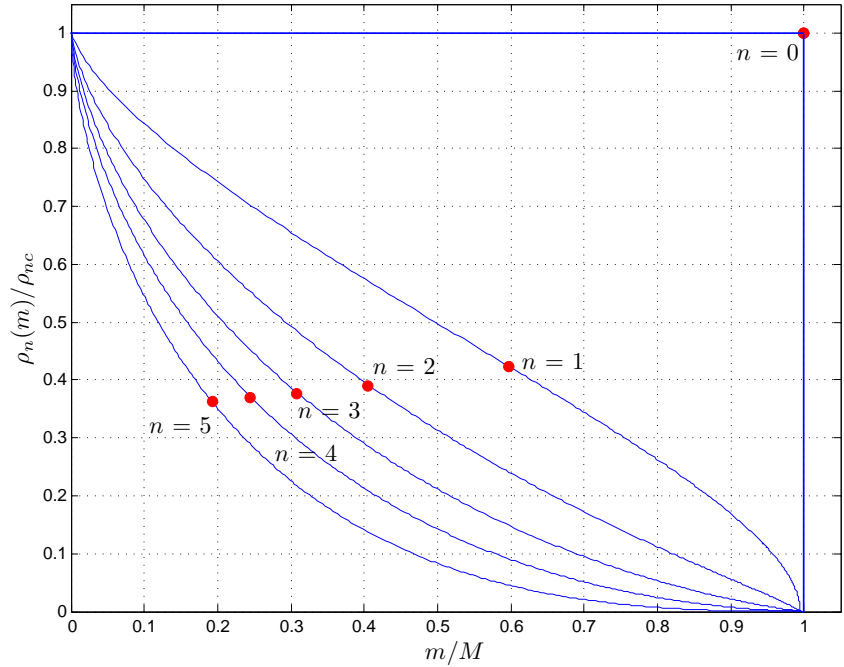}
\caption{Normalized density profiles as a function of fractional included mass $m/M$, for polytropes of finite mass $M$
and compressibility increasing with $n$. The red dots mark the cores. For incompressible matter ($n=0$), the polytrope
is all core. As the matter softens ($n$ increases), an envelope grows to ultimately encompass just over 80\% of the
mass. For any $n>1$, the density at the inner core radius stays in the narrow range
$0.37 < \rho(r_{{\rm ic}n})/\rho_c < 0.42$.}
\end{figure}

\begin{figure}[t] 
\includegraphics[scale=0.65]{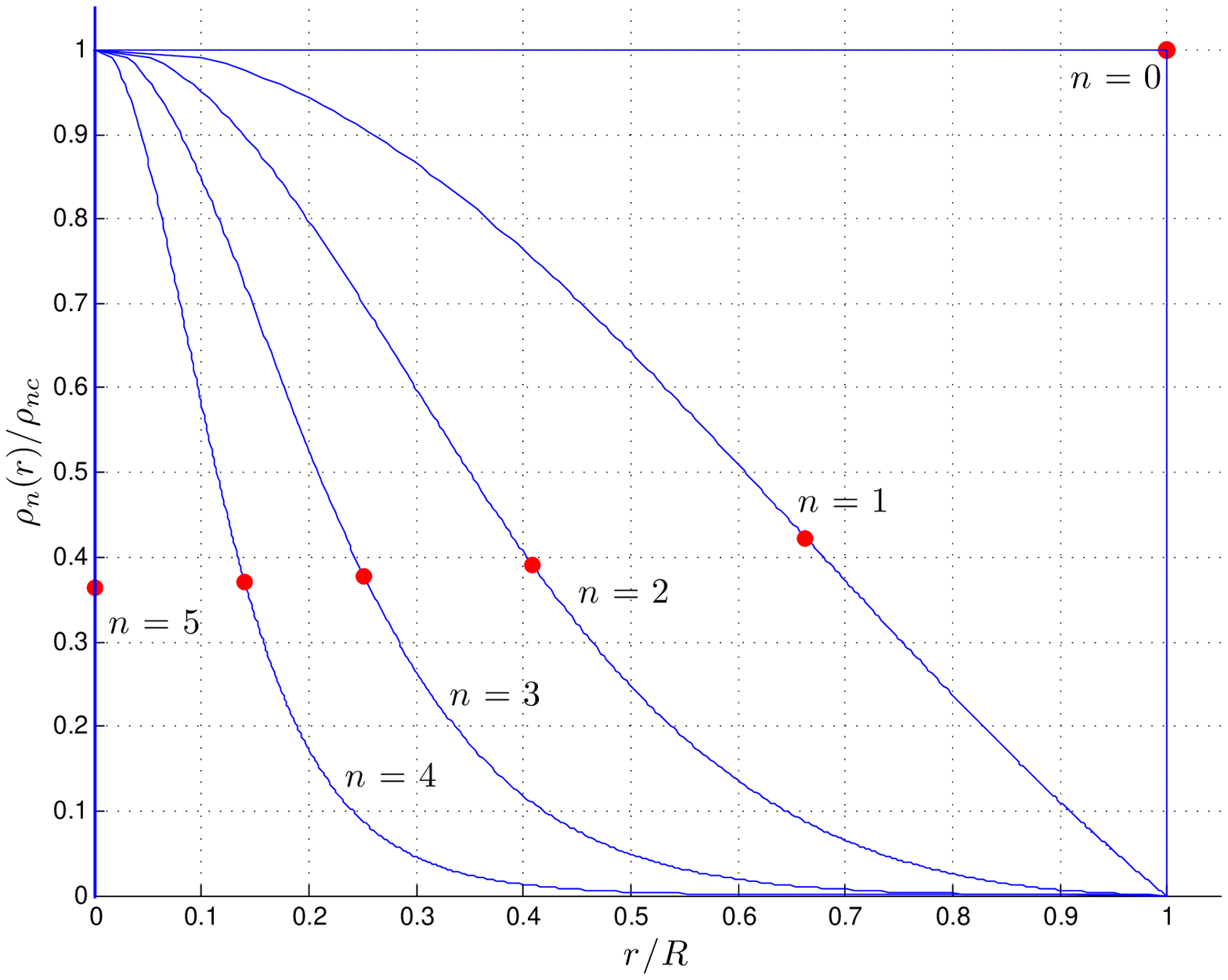}
\caption{Normalized density profiles as function of fractional radius $r/R$. The density is constant for incompressible
matter ($n=0$) which is all core, but is concentrated at the origin for an unbounded polytrope ($n=5,~R=\infty$)
which is all envelope.  For any $n>1$, the density at the inner core radius stays in the narrow range
$0.37 < \rho(r_{{\rm ic}n})/\rho_c < 0.42$.}
\end{figure}

For incompressible matter ($n=0$), there is no core concentration: the mass is uniformly distributed, and the entire
star is core.  But as the equation of state softens as $n$ increases toward $5$, the gradient
decreases, the core concentrates, the inner core radius shrinks, and the envelope outside the core grows:
$r_{{\rm ic}n}/R\rightarrow 0$, $m_{{\rm ic}n}/M\rightarrow \sim 0.19$.
For the softest equations of state $n\lesssim 5$, the stellar radius $\xi_{1n} \approx 3(n+1)/(5-n)$, the inner core
radius shrinks $\xi_{{\rm ic}n}\approx \sqrt{10/3 n}$, their ratio $r_{{\rm ic}n}/R=\xi_{{\rm ic}n}/\xi_{1n} \approx
0.045(5-n)$, $m_{{\rm ic}n}/M \approx 0.20$, and $_0\omega_n\approx\sqrt{3/\xi_{1n}}$.

For $n<5$, the stellar boundary lies at finite radius.  The Noether charge is nearly conserved at $G_n\approx 0$ in the inner core, but grows rapidly as the boundary is approached
(Figure~1). The polytropic form is locally scale invariant, but the radius $R$ determines the stellar scale.

For $n=5$, the core becomes infinitely concentrated, shrinking to zero, and
the star is all envelope.  The $n=5$ regular solution
\be \theta_5(\xi)=(1+\xi^2 /3)^{-1/2}\quad ,\quad \rho=\rho_c (1+\xi^2 /3)^{-5/2} \quad ,
\quad m=M\cdot\xi^3/(3+\xi^2)^{3/2}\quad , \quad v_5=(3-u)/3\quad .\ee
$_0\omega_5=0$, so that the $n=5$ polytrope has infinite stellar radius $R$ for any mass $M$ and is globally scale invariant. The Noether charge $G_5 \sim (v_5+u/3-1)=0$ everywhere.

\subsection{Lane-Emden Functions and Their Picard Approximations} 

For $0<n<5$,
\be w_n (z)=\int_0 ^z\ dz\ w_n\frac{(2-z+w_n /n)}{(3-z)(w_n-z)}\approx (5/J)[1-(1-z/3)^J]:=w_{n{\rm Pic}}(z)\quad ,\quad J:=(9n-10)/(7-n)\quad ,\ee
is well-approximated by the {\em Picard approximation}, obtained by inserting the core values $w_n (z)\approx$
$(5/3)z$ inside the integrals. Indeed, this Picard approximation is everywhere exact
for $n=0,~5$. For $0<n<5$,
it breaks down only in the outer envelope, where $w_n$ diverges as $w_n \rightarrow n[_0\omega_n ^{n-1}/u]^{1/n}$.
Integrating over $z$, the density profile and Lane-Emden functions are~\cite{BludKen}
\bea
\rho_n(z)/\rho_{cn}=\theta_n ^n =\exp{\Bigl\lbrace -\int_0 ^z  \frac{dz\ w_n(z)}{[w_n(z)-z](3-z)} \Bigr\rbrace} \approx
(1-z/3)^{5/2} \\
\theta_n = \exp{\Bigl\lbrace  -\int_0 ^z  \frac{dz\ w_n(z)}{n [w_n(z)-z](3-z)}\Bigr\rbrace} \approx
(1-z/3)^{5/2n}:=\theta_{n{\rm Pic}}\quad  ,\eea
where again the Picard approximations are obtained by inserting the core relations
$w_n(z)=n v_n(z) \approx (5/3)z$ inside the integrals.

\section{IN $P=K \rho^{4/3}$ POLYTROPES, RADIUS IS INDEPENDENT OF MASS}   

The $n=3$ polytrope, which is realized in relativistic degenerate stars (massive cold white dwarfs) and in the Eddington standard model (luminous ZAMS stars), is distinguished by a unique $M-R$ relation: the mass $M=4\pi ~_0\omega_3 ~(K/\pi G)^{3/2}$ depends only on
the equation of state constant $K:=P/\rho^{4/3}$, where the constant $K$ depends on the entropy but is independent of the radius $R$.  In these stars, the gravitational and internal energies cancel, making the
total energy $W=\Omega + U=0$. Such stars are in marginal dynamical equilibrium at any radius.

\subsection{Relativistic Degenerate Stars: $K$ Fixed By Fundamental Constants} 

Massive white dwarfs are supported by the degeneracy pressure of relativistic electrons, with number density $n_e=\rho/\mu_e m_H$, where $m_H$ is the atomic mass unit and the number of electrons per atom $\mu_e=Z/A=2$, because massive white dwarfs are composed of pure He or $C^{12}/O^{16}$ mixtures. Consequently, $K_{WD}=\frac{hc}{8} [\frac{3}{\pi}]^{1/3} {m_H \mu_e}^{-4/3}$ depends only on fundamental constants. This universal value of $K_{WD}$ leads to the limiting Chandrasekhar mass $M_{Ch}=\frac{\pi^2}{8 \sqrt{15}} M_{\star}/\mu_e^2=5.83 M_{\odot}/\mu_e ^2=1.46 M_{\odot}$.

\subsection{Zero-Age Main Sequence Stars: $K(M)$ Depends on Specific Radiation Entropy} 

In a non-degenerate ideal gas supported by both gas pressure $P_{\rm gas}=\mathcal{R}\rho T/\mu$ and radiation pressure $P_{\rm rad}=a T^4 /3$, the total pressure $P=P_{\rm gas}+P_{\rm rad}=P_{\rm gas}[1+(1-\beta )/\beta ]=P_{\rm gas}[1+ \mu S_{\rm rad}/4\mathcal{R}]$, where $\beta:=P_{\rm gas} /P$ and
\be S_{\rm rad}=\frac{4a T^3}{\rho}=(\mathcal{R}/\mu)\cdot 4 (1-\beta)/\beta\quad ,\quad S_{\rm gas} =(\mathcal{R}/\mu)\cdot \log{(T^{3/2}/\rho)} \ee
are the specific radiation entropy and gas entropy for an ideal monatomic gas.
Bound in an $n=3$ polytrope, the gas entropy is a constant depending on $K$, the total specific entropy
is constant, and the total energy $W=\Omega + U$ vanishes.  Because such stars are in marginal dynamical equilibrium at any radius, their mass $M$ is independent of radius $R$.

In zero-age Main Sequence (ZAMS) stars of mass $M$, the radiation entropy $S_{\rm rad}$ is approximately constant,
because at each radius, radiation transport leaves the luminosity generated by interior nuclear energy generation proportional to the local transparency (inverse opacity) $\kappa ^{-1}$.  {\em Eddington's standard model} assumes constant  $S_{\rm rad} (M)$. This makes $\beta (M)$ and $K(M)=\lbrace [3(1-\beta)/a] (\mathcal{R}/\mu\beta )^4 \rbrace^{1/3}$ constants depending only on $M$, according to {\em Eddington's quartic equation}
\be \frac{1-\beta}{\beta ^4}=\Bigl( \frac{M\mu^2}{M_{\star}} \Bigr)^2\quad ,\quad M_{\star}:=\frac{3\sqrt{10} ~_0\omega_3}{\pi ^3} \Bigl( \frac{hc}{G m_H^{4/3}} \Bigr)^{3/2}=18.3 M_{\odot}\quad .\ee

The luminosity $L=L_{Edd} [1-\beta (M)]$ depends on the {\em Eddington luminosity} $L_{Edd}:=4\pi c G M/\kappa_p$ through the photospheric opacity $\kappa_p$. From Eddington's quartic formula, the stellar luminosity
\be L/L_{\odot}=\frac{4 \pi c G M_\odot}{\kappa_s L_\odot} (0.003)\mu ^4 \beta  ^4 (M/M_\odot)^3 .\ee
This is confirmed~\cite{Hansen} in ZAMS stars, particularly over the lower end of the Main Sequence $0.1 M_{\bigodot} <M<100 M_{\bigodot}$.

\begin{table*}[t] 
\caption{Taylor Series and Picard Approximations $\theta_{n{\rm Pic}}$ to Lane-Emden Functions $\theta_{n}$}
\begin{ruledtabular}
\begin{tabular}{|l||l||l|l}
$n$  &Lane-Emden Function and Taylor Series &$N:=5/(3n-5)$ &Picard Approximation $\theta_{n{\rm Pic}}:=
(1+\xi^2/6N)^{-N}$  \\
\hline \hline
0    &$1-\xi^2/6$                                   &-1         &$1-\xi^2/6$                              \\
1    &$\sin{\xi}/\xi=1-\xi^2/6+\xi^4/120-\xi^6/5040+\cdots$&-5/2&$(1-\xi^2 /15)^{5/2}=1-\xi^2/6+\xi^4/120-\xi^6/10800+\cdots$    \\
$n$  &$1-\xi^2/6+n \xi^4/120-n(8n-5)/15120 \xi^6+\cdots$   &$5/(3n-5)$ &$(1+\xi^2/6N)^{-N}=1-\xi^2/6+n \xi^4/120-n(6n-5) \xi^6 /10800+\cdots$ \\
5    &$(1+\xi^2/3)^{-1/2}$                                 &1/2        &$(1+\xi^2/3)^{-1/2}$ \\
\end{tabular}
\end{ruledtabular}
\end{table*}

\section{ANALYTIC APPROXIMATIONS TO LANE-EMDEN FUNCTIONS} 
We now find analytic approximations to the radial structure of any polytrope, particularly the important $n=3$ polytrope.
From the solutions $w_n(z)$ to the first-order equation, we now use
\be dm/m:=u\cdot dr/r=dz/[w_n(z)-z] \label{eq:chareqns} \ee
to obtain
\bea
m(z)/M=(\frac{z}{3})^{3/2}\cdot \exp{\Bigl\lbrace \int_0 ^z dz \Bigl\lbrace
\frac{1}{[w_n(z)-z]}-\frac{3}{2z}\Bigr\rbrace \Bigr\rbrace} \approx (\frac{z}{3})^{3/2}\\
r(z)/R=(\frac{z}{3})^{1/2}\cdot \exp{\Bigl\lbrace  \int_0 ^z  dz \Bigl\lbrace \frac{1}{(3-z)[w_n(z)-z]}-
\frac{1}{2z}\Bigr\rbrace \Bigr\rbrace } \approx \frac{(3z)^{1/2}}{3-z}\quad  \eea
for the mass and radial distributions.
The integration constants $R,~M,~\rho_c$ express the scale dependence of the polytrope.
Using equation (\ref{eq:chareqns}) to eliminate $z(\xi)$, the Picard approximations
\be \theta_{n{\rm Pic}}(\xi)=(1+\xi^2/6N)^{-N}\quad, \quad N:=5/(3n-5) \ee
to the Lane-Emden functions are obtained and tabulated
in the last column of Table~I. For $n=0$ and 5 polytropes, this closed form is exact.
For intermediate polytropic indices $0<n<5$, the Picard approximation breaks down near the outer boundary, but remains a good approximation over most of the polytrope's bulk. The worst Picard approximation is for $n=3$.

For $n=3$, the Taylor series expansion
\bea \theta_3(\xi)= 1-\xi^2/6+\xi^4 /40-(19/5040) \xi^6+(619/1088640) \xi^8 -(2743/39916800) \xi^{10} + \cdots \\ =1-0.1666667 \xi^2+ 0.025 \xi^4 - 0.0037698 \xi^6 +  0.0005686 \xi^8-0.00006872 \xi^{10} + \cdots,  \eea
when truncated at tenth-order, has a radius of convergence at $\xi \sim 2.4$. Outside this radius,
the Picard approximation (15)
\be \theta_{3{\rm Pic}}(\xi)=\theta_3 +\xi^6 /3528 + \cdots \ee
overestimates the Taylor series expansion, but remains a good approximation over the bulk of the star, with error $\leq 10\%$ out to $\xi\approx 3.9$, more than twice the core radius and more than half-way out to the stellar boundary at $\xi_{13}=6.897$. The Picard approximations in white dwarf and ZAMS stars suffices, except for their very outer envelopes, which contain little mass and are never polytropic.

Because it satisfies the central boundary condition but not the outer boundary condition, the Picard approximation underestimates $\theta ' (\xi)$ and overestimates $\theta (\xi)$ outside $\xi \sim 3.9$.  A much better and simpler approximation is the {\em Padé rational approximation}~\cite{Seidov}
\be \theta_{\rm 3Pad}:=
\frac{1-\xi^2/108+11 \xi^4/45360}{1+119 \xi^2/756+ \xi^4/1008}=1. - 0.166667 \xi^2 + 0.025 \xi^4 - 0.00376984 \xi^6 +
 0.0005686 \xi^8 - 0.0000857618 \xi^{10}+\cdots , \ee which, already in tenth-order, starts converging faster than the Taylor series. In fact, this Pad\'{e} approximation agrees almost exactly with the exact solution out to
$\xi_1=6.921$, very close to $\xi_{13}=6.897$, the first zero of the exact Lane-Emden function $\theta_3(\xi)$.
These simple analytic approximations to $\theta_3 (\xi)$, shown in Figure~5, simplify structural modeling of massive white dwarfs and ZAMS stars.

\begin{figure}[t] 
\includegraphics[scale=0.60]{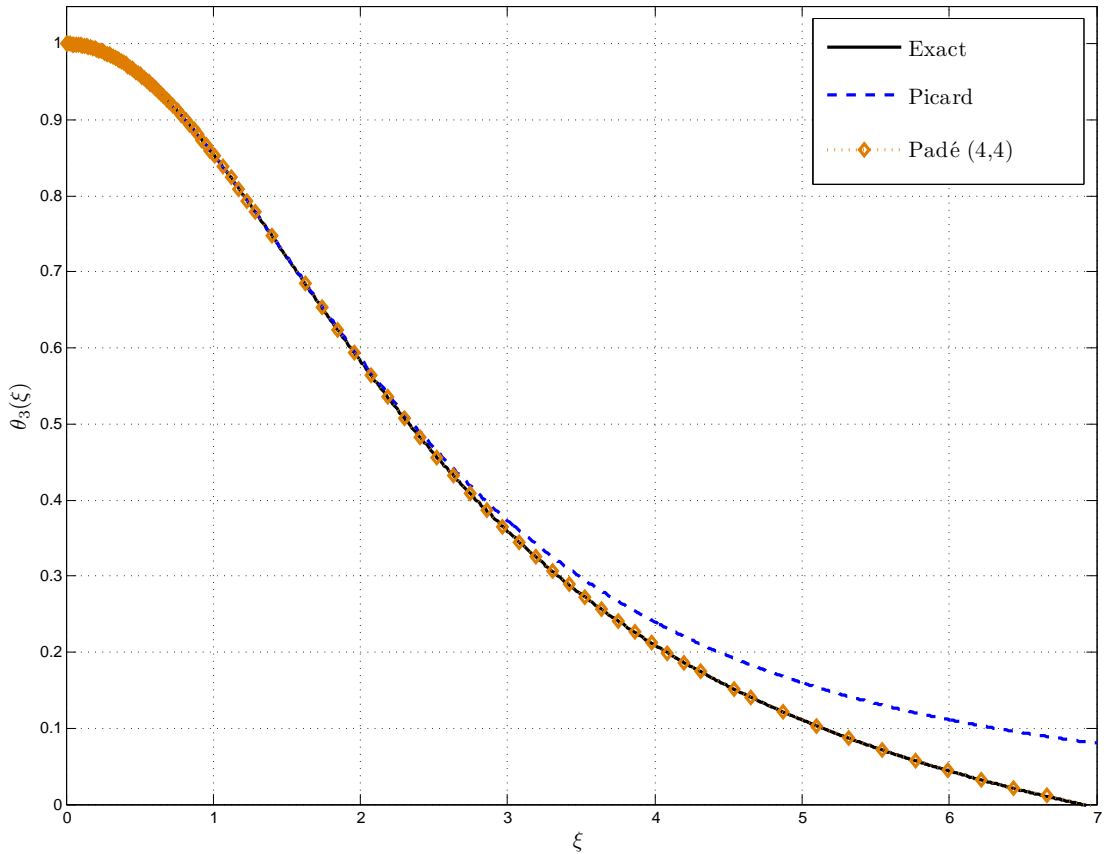}
\caption{The exact Eddington Standard Model Lane-Emden function $\theta_3(\xi)$ and its Picard and Pad\'{e}
approximations. Even in this worst case, the Picard approximation works well out to twice the core radius at
$\xi_{{\rm ic}3}=\xi_{13}(r_{{\rm ic}3}/R)$ = 1.65, but breaks down near the boundary. The Pad\'{e} approximation is indistinguishable from the exact solution, vanishing at $\xi_1=6.921$, very close to the exact zero $\xi_{13}=6.897$ (For $n\neq 3$ polytropes, the Picard approximations would be even better and become exact everywhere as $n \rightarrow$ 0 or 5.) .}
\end{figure}

\section{CONCLUSIONS}

For spherical hydrostatic systems obeying barotropic equations of state,
the scaling evolutionary equation leads to a linear relation between the local gravitational and internal
energies.  From this equation, directly follow all the properties of polytropes and the important mass-radius relation. Quadrature then leads to the
regular Lane-Emden functions and their Picard and Padé approximations, which are useful wherever stars are approximately or exactly polytropic.

We illustrated this for the $n=3$ regular polytrope and obtained analytic approximations to the solution of the Lane-Emden equation, valid over the bulk of massive white dwarfs and ZAMS stars in radiative equilibrium.
\begin{acknowledgments}
SAB thanks Romualdo Tabensky (Universidad de Chile) for helpful discussions of Picard and Pad\'{e} approximations and acknowledges support from the Millennium Center for Supernova Science through grant P06-045-F funded
by Programa Bicentenario de Ciencia y Tecnolog\'ia de CONICYT and Programa Iniciativa Cient\'ifica Milenio de
MIDEPLAN. The figures were generated with MATLAB~7.
\end{acknowledgments}

\bibliography{bibliographyLE}
\end{document}